\documentclass[epj]{svjour}
\usepackage{graphics}
\begin{document}
\title{Coevolution of competing systems: local cooperation and global inhibition}
\author{Jos\'e M. Albornoz\inst{1,2} \and Antonio Parravano\inst{2}
}                     
%
%
\institute{Departamento de Electr\'onica y Comunicaciones, Facultad de Ingenier\'ia,
Universidad de Los Andes, La~Hechicera, M\'erida, M\'erida~5251, Venezuela. \and Centro de F\'{\i}sica Fundamental,
Facultad de Ciencias, Universidad de Los Andes, Apartado Postal 26, La~Hechicera, M\'erida, M\'erida~5251, Venezuela.}
\date{Received: date / Revised version: date}
%
\abstract{
Using a set of heterogeneous competing systems with intra-system cooperation and inter-system aggression, we show how the coevolution of the system parameters (degree of organization and conditions for aggression) depends on the rate of supply of resources $\dot{S}$. The model consists of a number of units grouped into systems that compete for the resource $S$; within each system several units can be aggregated into cooperative arrangements whose size is a measure of the degree of organization in the system. Aggression takes place when the systems release inhibitors that impair the performance of other systems. 
Using a mean field approximation we show that i) even in the case of identical systems there are stable inhomogeneous solutions, ii) a system steadily producing inhibitors needs large perturbations to leave this regime, and iii) aggression may give comparative advantages. A discrete model is used in order to examine how the particular configuration of the units within a system determines its performance in the presence of aggression. We find that full-scale, one sided aggression is only profitable for less-organized systems, and that systems with a mixture of degrees of organization exhibit robustness against aggression. By using a genetic algorithm we find that, in terms of the full-occupation resource supply rate $\dot{S}_{F}$, the coevolution of the set of systems displays the following behavior: i) for $\dot{S}< \dot{S}_{F}/10$ aggressions are irrelevant and most systems exhibit a high degree of organization; ii) For $\dot{S}_{F}/10 < \dot{S} < \dot{S}_{F}/3$ aggressions are frequent, making systems with a low degree of organization competitive; iii) for $\dot{S}_{F}/3 < \dot{S} < \dot{S}_{F}/2$  the systems display global evolutive transitions between periods of calm (few aggressions and high degree of organization) and periods of belligerence (frequent aggressions and low degree of organization); iv) for $ \dot{S} > \dot{S}_{F}/2$ the periods of aggression becomes progressively rarer and shorter. Finally, when $\dot{S}$ approaches $\dot{S}_{F}$ the selection pressure on the cooperativity and the aggression between systems disappears. This kind of model can be useful to analyse the interplay of the cooperation/competition processes that can be found in some social, economic, ecological and biochemical systems; as an illustration we refer to the competition between drug-selling gangs. 
} 

\def\beq{\begin{equation}}
\def\eeq{\end{equation}}
\def\av{\langle \upsilon \rangle}
\def\avi{\langle \upsilon_i \rangle}
\def\avm{\langle \upsilon \rangle _{M}}
\def\avd{\langle \upsilon \rangle _{D}}
\def\avt{\langle \upsilon \rangle _{T}}
\def\avmdt{\langle \upsilon \rangle _{MDT}}
\def\barvi{\overline{\upsilon}_i}
\def\vcri{\overline{\upsilon}_{cri}}
\def\barfm{\overline{f}_M}
\def\barfbot{\overline{f}_{M,bot5}}
\def\barftop{\overline{f}_{M,top5}}

\maketitle
\section{Introduction}
\label{intro}
There are many situations in which the global performance of a set of competing systems, measured in some way, is below its maximum possible value due to competition among the systems. Generally, in complex systems the competing components use different strategies that may consume part of the available resources in order to reduce the performance of the others. In these situations there is a global cost but the collateral effect is to favor the emergence of new strategies that tend to increase the diversity and complexity in these systems. This is the case in some social and economic systems \cite{Axelrod86,Axelrod97,Cara2000,kuper02,szabo04} and in certain ecological and biochemical systems \cite{frank95,Pfeiffer2001,MacLean06,bra01,hsu04}. For example, Axelrod \cite{Axelrod86,Axelrod97} considers an evolutionary approach to social norms based on a n-person Prisoner's dilemma. In this  game the players can defect (getting a payoff and hurting each of the other players) but can be punished if seen by another player. The evolution of the player's strategy (the boldness and the vengefulness) is driven by the selection of the strategies giving the best scores; however, the maximum global score (no one defects) is rarely observed.

There are several instances in which there is an interplay between risk and profit. In some cases, the balance between risk and profit depends on the degree of organization within a competing system: a greater degree of organization usually results in a greater profit but on the other hand it may also increase potential losses in case of an aggression from a competing system. Consider for instance the various drug-selling gangs that coexist in the poorest neighborhoods of a large city \cite{Block93,Levitt00}. 
The gangs compete to sell drugs to a limited number of customers (the resource). The customers prefer to buy in the safest, easiest and quickest possible way and the gangs try to sell as much as they can by using strategies to attract the largest possible fraction of the consumers in the city. The model presented here considers two types of strategies: intra-system cooperation and inter-system aggression. In the gangs example, the degree of intra-gang cooperation can be associated to the various ways in which the gang members are grouped. The dealers can distribute themselves in as many intersections as dealers in the gang, however busy corners (as well as shopping centers) where several dealers operate at the same time offer an environment in which customers can buy drugs in an easy and quick way; besides, the presence of other customers gives the impression of a relatively safe place, hence increasing the probability of a successful transaction. Inter-system aggressions can be associated with the violence between gangs. In our model an aggression has a cost for the aggressor and causes more damage to units working in cooperation than to isolated units. The shooting of a rival drug-dealer immediately stops drug sales at the crime scene. During the next days or weeks the police is patrolling the area, scaring away habitual customers who choose to buy drugs in safer places. If the shooting occurs in a busy corner, where several drug-dealers operate simultaneously, income losses per shooting are larger compared to the one-dealer one-corner case. On the other hand, the attacking gang pays a cost in several ways: money to buy guns and/or hire mercenary aggressors, demand of dealers for higher wages because of the added risk, etc. This particular example illustrates the kind of competition/cooperation mechanisms considered in this paper that are common to certain socioeconomic, ecological, and biochemical systems. Our analysis considers these general aspects; however we will refer to the drug-gang example in order to
 illustrate the relation between our general (non specific) model and a concrete, real-world situation.

The aim of this work is to formulate a model to analyse the coevolution of the strategies of systems competing for limited resources \cite{pugliese09}; specifically, we study the coevolution of local intra-system cooperation and global inter-system aggression. Our model consist of $N_g \times N_e $ identical units organized in $N_g$ systems with $N_e$ elements per system. This type of systems is found in cells, where enzymes are grouped within specialized organelles; similarly, in the industrial sector, machines are grouped in factories, or, as in the gang example, drug dealers are grouped in gangs. The systems compete to acquire the resources (substrates, raw material or customers) supplied to the system at a rate $\dot S$ and at the same time are allowed to produce inhibitors that reduce the performance of others. The cooperation among associated units increases their efficiency to acquire the resources when they are arranged in oligomers, as in the case of enzymes \cite{segel75}; the number of units cooperating in a given configuration can be interpreted as a measure of its degree of organization. Aggression is modelled by the release of inhibitors, which reduce the efficiency by blocking the unit during a period of time as in the case of the occupation by an inhibitor of the active site where a substrate is bound to an enzyme \cite{segel75}. The performance (i.e. the production rate) of a system depends on its comparative ability to acquire the resources, which in turn depends on the configuration of the units in the system. In absence of inhibitors, the production rate is greater in systems where units cooperate than in systems where units work in isolation. However, when inhibitors are present, the number of units blocked per inhibitor depends on the working configuration of the units.  In this way, the effect of an aggression will be more damaging to systems with a greater degree of organization (greater number of units working in cooperation). 

In real systems the strategies used by its components to survive and reproduce are in general very sophisticated and are the result of evolution. Models like the one presented here help to understand how environmental conditions and interaction between the systems drive the evolution of the model parameters; in our case, the fraction of units working in isolation and the conditions that must be satisfied for the release of inhibitors. 

The model is presented in Sec. \ref{sec:2}, both in a continuous as well as in a discrete version. In Sec. \ref{sec:3} we consider a set of competing systems including four types of systems with different cooperating configurations; two cases are considered: when the inhibitors are introduced by an external source and when the inhibitors are released by the systems. In Sec. \ref{sec:4} a genetic algorithm is used to allow the evolution of the systems' strategies when the units operate either in isolation or in cooperative arrangements of four units each. Conclusions are given in Sec. \ref{sec:5}.

\section{The model}
\label{sec:2}
Consider a set of $N_g$ systems, each one with the same number $N_e$ of units. 
The $N_g \times N_e$ units compete to acquire the available resources (hereafter known as the substrate) that are supplied at a rate $\dot{S}$. Each unit can be in one of four states: i) idle (ready to acquire either a substrate or an inhibitor), ii) busy (processing a substrate during a period of time $\tau_p$), iii) recovering (during a period of time $\tau_r$ necessary to reach the idle state after releasing a product), or iv) inhibited (unable to acquire a substrate during a period of time $\tau_I$ after acquiring an inhibitor). Therefore, the duty cycle of a non-inhibited unit is $\tau = \tau_p + \tau_r$. The units in a system can be organized in cooperative arrangements of $\mu$ units; such arrangements increase the efficiency of the units to acquire either substrates or inhibitors. When $\kappa$ of the $\mu$ units in a cooperative arrangement are busy the remaining $\mu-\kappa$ idle units increase their probability to bind to a substrate (or to a inhibitor) by a factor $\alpha^{-\kappa}$ relative to the isolated case, where $1/\alpha$ ($0<\alpha<1$) is the efficiency of the cooperation. If a unit is inhibited, the other idle units in the cooperative arrangement are also inhibited.

In the following two subsections we present a continuous and a discrete version of the model. In the continuous version (Sec. \ref{subsec:2.1}) we use a mean field approach to obtain the stationary solutions for the average state of the $N_g$ systems. 
These stationary solutions are useful to visualize the main properties of the model, however they cannot be used to follow the dynamics of the individual units nor to describe the general case of inhomogeneous intra-system cooperativity. The discrete version of the model (Sec. \ref{subsec:2.2}) allows to follow the evolution of the state of each unit, of each cooperative arrangement, and the production rate of each system. It must be borne in mind that the continuous model presented here does not represent all the details contained in the discrete model; nevertheless its stationary solutions shed light on the essential properties of the type of systems we want to study.

\subsection{Mean Field approximation: stationary solutions}
\label{subsec:2.1}
In order to show some of the basic properties of the model, we find the stationary solutions when the internal structure of each system is described by the fraction of idle units $F_i$ in system $i$ ($1\leq i \leq N_g$). The recovering state of each unit is neglected so that the duty cycle is $\tau=\tau_p$.
We approximate the total production rate $\dot{G}_i$ of products $P$ plus inhibitors $I$ in system $i$ as
\begin{equation}
\dot{G}_i=\dot{P}_i+\dot{I}_i= N_e F_i p_i S,
\label{stat:1}
\end{equation}
where $S$ is the stationary number of substrates that can be bound by the $N_g \times N_e$ units, 
$N_e F_i$ is number of idle units in system $i$, and
\begin{equation}
p_i = \frac{p_0}{\alpha^{\langle\kappa\rangle_i}} \simeq \frac{p_0}{\alpha^{(\mu_i-1)(1-F_i)}}
\label{stat:2}
\end{equation}
is the probability (per unit time) that an idle unit binds either to a substrate or to an inhibitor. In this expression the parameter $p_0$ is the probability (per unit substrate and unit time) that an isolated idle unit binds to a substrate (or to an inhibitor), $\alpha$ controls the degree of cooperation between the components of a cooperative arrangement, $\mu_i$ is the number of units in the cooperative arrangements of system $i$, and
$\langle\kappa\rangle_i \simeq (\mu_i-1)(1-F_i)$ approximates the average number of busy units in a cooperative arrangement in system $i$. 
Note that the number of units $\mu$ per cooperative arrangement is assumed to be constant in a given system, but can differ from system to system.
These approximations are overcome in the discrete model (Sec. \ref{subsec:2.2}).

The stationary fraction of occupied units $F_{occ,i}$ and inhibited units $F_{inh,i}$ in system $i$ is approximated as
 \begin{equation}
F_{occ,i}= S F_i p_i \tau \,\,\,\,\, {\rm and} \,\,\,\,\,  F_{inh,i}= I F_i p_i \mu_i \tau_{I,i}.
\label{stat:3}
\end{equation}
where $\tau_{I,i}$ is the time during which a cooperative arrangement in system $i$ is disabled once one of its units binds an inhibitor, and $I$ is the stationary number of inhibitors available to be bound.
Since $F=1-F_{occ}-F_{inh}$, the equation
\begin{equation}
\frac{1-F_i}{F_i} \frac{\alpha^{(\mu_i-1)(1-F_i)}}{p_0}= S \tau + I \mu_i \tau_{I,i}
\label{stat:4}
\end{equation}
provides $F_i$ as a function of $S$, $I$, and the model parameters $\alpha$, $p_0$, $\mu_i$, $\tau_i$, and $\tau_{I,i}$. 
In stationary conditions
\begin{equation}
S=\frac{\dot{S}}{N_e \sum p_i F_i}
\label{stat:5}
\end{equation}
and 
\begin{equation}
I=\frac{\sum \dot{I}_i}{N_e \sum p_i F_i}.
\label{stat:6}
\end{equation}

\begin{figure}
\resizebox{0.45\textwidth}{!}{%
  \includegraphics{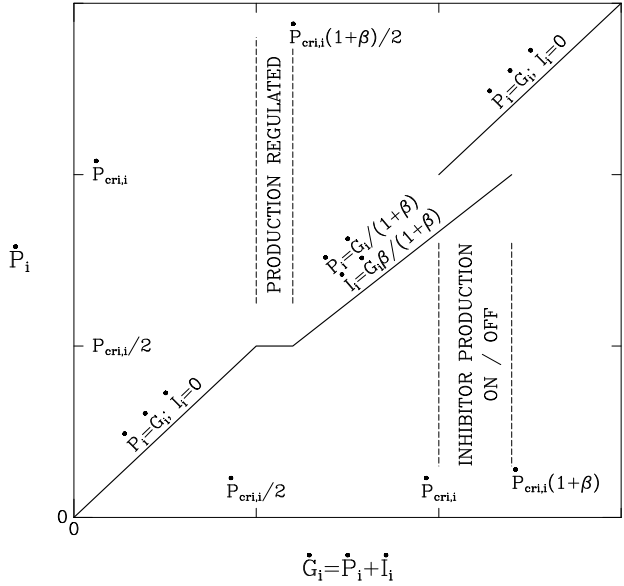}
}
\caption{The production rate $\dot{P}_i$ as function of the total rate production $\dot{G_i}=\dot{P}_i+\dot{I}_i$. The inhibitor production is $\dot{I}_i=\beta \dot{P}_i$ for $\dot{P}_{cri,i}/2< \dot{P}_i<\dot{P}_{cri,i}$, whereas $\dot{I}_i=0$ for 
$\dot{P}_i< \dot{P}_{cri,i}/2$ and $\dot{P}_i> \dot{P}_{cri,i}$. For $\dot{P}_{cri,i}/2<\dot{G_i}<(1+\beta)\dot{P}_{cri,i}/2$ the production rate is regulated to the value $\dot{P}_{cri,i}/2$ and the inhibitor production is $\dot{I}_i=\dot{G_i}-\dot{P}_{cri,i}/2$. In the range $\dot{P}_{cri,i}<\dot{G_i}<(1+\beta)\dot{P}_{cri,i}$ the inhibitor production is bi-valued.
}
\label{fig:1}     
\end{figure}

Finally, the rate of inhibitor production is assumed to depend on the total production rate $\dot{G_i}=\dot{P}_i+\dot{I}_i$ as shown in Fig. \ref{fig:1}. 
Inhibitors are produced only when the production rate is in the range 
$\dot{P}_{cri,i}/2<\dot{P}_i<\dot{P}_{cri,i}$, where $\dot{P}_{cri,i}$ is the critical production rate below which system $i$ releases inhibitors. Since an aggression has an associated cost, a system must achieve a minimum performance level in order to release an inhibitor; therefore inhibitors will not be released if $\dot{P}_i  < \dot{P}_{cri,i}/2$. Within the $\dot{P}_{cri,i}/2<\dot{P}_i<\dot{P}_{cri,i}$ range the rate of inhibitor production is proportional to $\dot{P}$; that is, $\dot{I}=\beta \dot{P}$.
Note that:
i) In the range $\dot{P}_{cri,i}/2<\dot{G_i}<(1+\beta)\dot{P}_{cri,i}/2$
the production rate is regulated to the value $\dot{P}_i=\dot{P}_{cri,i}/2$; that is, within this range any change in $\dot{S}$ will produce a change in $\dot{I}_i$, but not in $\dot{P}_i$.
ii) In the range $\dot{P}_{cri,i}<\dot{G_i}<(1+\beta)\dot{P}_{cri,i}$ the system adopts one of two distinct regimes: one in which $\dot{P_i}=\dot{G_i}$ and $\dot{I}_i=0$, and the other in which 
$\dot{P_i}=\dot{G_i}/(1+\beta)$ and $\dot{I}_i=\dot{G_i} \, \beta/(1+\beta)$. These two regimes result in a hysteresis behavior, as follows: a) starting in the upper branch and decreasing $\dot{S}$, no inhibitors are produced until the production rate reaches the critical value $\dot{P}_{cri,i}$, b) starting in the lower branch and increasing $\dot{S}$, the system produces inhibitors until the production rate reaches the critical value $(1+\beta) \, \dot{P}_{cri,i}$.

We are now interested in the stationary solutions $\dot{P}_i(\dot{S})$ and $\dot{I}_i(\dot{S})$. An iterative procedure is employed to obtain them: starting with an initial guess value for $I$, the value of $F_i$ in each system is obtained through eqs. (\ref{stat:4}) and (\ref{stat:5}), while $\dot{G_i}$ is obtained by using eq. (\ref{stat:1}); then function $\dot{G_i}(\dot{P}_i)$ is used to obtain new values for $\dot{I}_i$. Equation (\ref{stat:6}) is used to obtain a new value of $I$, and the procedure is repeated until convergence is achieved. To deal with the bi-valued region of $\dot{G_i}(\dot{P}_i)$ the procedure is repeated considering all possible combinations of the systems' behavior (i.e. the inhibitor production being ON or OFF in the bi-valued region).

\begin{figure}
\resizebox{0.45\textwidth}{!}{%
  \includegraphics{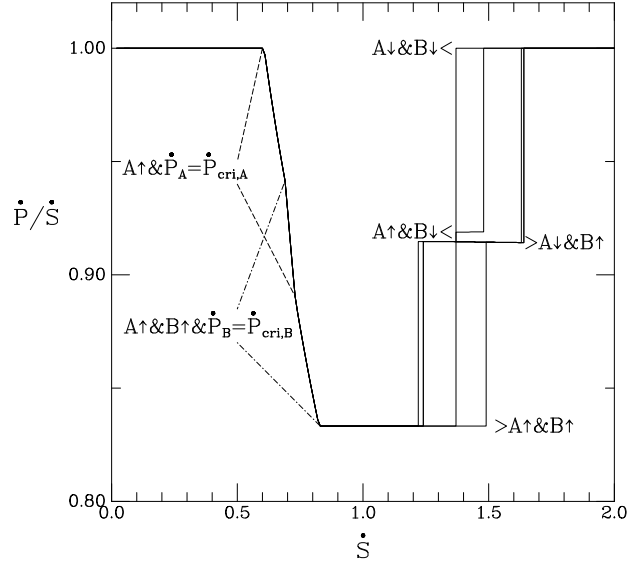}
}
\caption{Normalized total production rate $\dot{P}/\dot{S}=(\dot{P}_A+\dot{P}_B)/\dot{S}$ as a function of the supply rate of substrate $\dot{S}$ for two systems labeled A and B. 
The two systems are identical, except for the size of cooperative arrangements $\mu$ and the critical production rate $\dot{P}_{cri}$, which are are greater in system B.
The parameter values are $\beta=0.2$, $p_0=0.1$, $\alpha=0.5$, $\tau_A=\tau_B=1$, $\tau_{I,A}=\tau_{I,B}=5$, $\mu_A=1$, $\mu_B=2$, $\dot{P}_{cri,A}=0.6$, and $\dot{P}_{cri,B}=0.7$.
}
\label{fig:2}
\end{figure}

Figure \ref{fig:2} shows the equilibrium solutions in the plane [$\dot{P}/\dot{S}$, $\dot{S}$] for two systems, labeled A and B. For the chosen parameter values, the production regulated regimes of the two systems partially overlap, as well as the bi-valued regimes. Note that all possible combinations occur: the symbol A$\uparrow$ (A$\downarrow$) indicates that system A is producing (not producing) inhibitors. Even two identical systems (for instance, the case considered in Fig. \ref{fig:2} but with $\mu_A=\mu_B$ and
$\dot{P}_{cri,A}=\dot{P}_{cri,B}$) can display heterogeneous behavior in the range $\dot{P}_{cri}<\dot{G}<(1+\beta)\dot{P}_{cri}$. In this range two homogeneous (A$\uparrow$\&B$\uparrow$ and A$\downarrow$\&B$\downarrow$) and two inhomogeneous (A$\uparrow$\&B$\downarrow$ and A$\downarrow$\&B$\uparrow$) solutions can be distinguished in Fig. \ref{fig:2}. This behavior is a consequence of the hysteresis in $\dot{G_i}(\dot{P}_i)$, which in turn is characteristic of real systems with competition, cooperation, and aggression: once a conflict arises it tends to persist over time, and a substantial change in the conditions is required to extinguish it \cite{coleman07}.

We now focus in a particular case in which system A is constituted by single, non cooperative units (i.e. $\mu_A=1$) with the capacity to produce inhibitors ($\dot{I}_A=\beta \dot{P}_A$), whereas system B is composed by cooperative arrangements (i.e. $\mu_B>1$) that never produce inhibitors ($\dot{I}_B=0$). In this case, the fraction of idle units in the systems satisfy the equations
\begin{equation}
\frac{1-F_A}{F_A}= \frac{\dot{S}}{N_e \sigma} [\tau + \frac{\tau_I \, F_A \, \beta}{\sigma (1+\beta)}],
\label{stat:7}
\end{equation}
and
\begin{equation}
\frac{1-F_B}{F_B} \alpha^{(\mu_B-1)(1-F_B)}= \frac{\dot{S}}{N_e \sigma} [\tau + \frac{\mu_B \,  \tau_I\,  F_A \, \beta}{\sigma (1+\beta)}],
\label{stat:8}
\end{equation}
where
\begin{equation}
\sigma= F_A + \frac{F_B}{\alpha^{(\mu_B-1)(1-F_B)}}.
\label{stat:9}
\end{equation}

$F_A$ and $F_B$ can be obtained by solving eqs. (\ref{stat:7}) and (\ref{stat:8}) for given parameters $\dot{S}$, $\beta$, $\alpha$, $\mu_B$, $\tau$ and $\tau_I$. The production rates are then given by
\begin{equation}
\dot{P}_A=\frac{\dot{S}\,F_A}{\sigma(1+\beta)} \,\,\,\,\,\, {\rm and} \,\,\,\,\,\,  \dot{P}_B=\frac{\dot{S}\,F_B}{\sigma \, \alpha^{(\mu_B-1)(1-F_B)}}.
\label{stat:10}
\end{equation}

\begin{figure}
\resizebox{0.45\textwidth}{!}{%
  \includegraphics{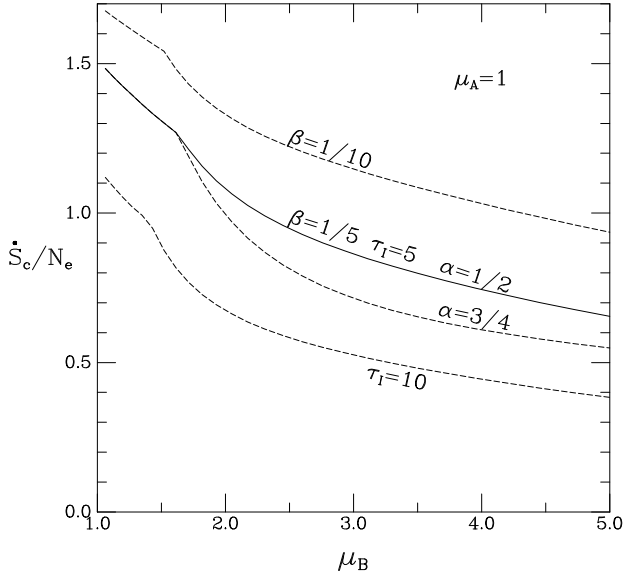}
}
\caption{
The normalized critical value $\dot{S}_c/N_e$ of the substrate input rate at which $\dot{P}_A=\dot{P}_B$ when system A converts a fraction $\beta$ of its products into inhibitors and its units do not cooperate (i.e. $\mu_A=1$). For substrate supply rates greater than $\dot{S}_c$ the production rate of system A is larger than that of system B. 
The continuous curve corresponds to parameter values $\beta=0.2$, $p_0=0.1$, $\alpha=0.5$, $\tau_A=\tau_B=1$, $\tau_{I,A}=\tau_{I,B}=5$ and $\mu_A=1$. The dashed curves shows the effect of varying $\beta$, $\tau_I$ or $\alpha$.
}
\label{fig:3}
\end{figure}

System A always uses a fraction $\beta$ of the processed substrate to release inhibitors (thus reducing its production efficiency), whereas system B converts all the processed substrate into products. Moreover, the idle units in system B are more efficient at acquiring substrates ($\mu_B>1$ and $\alpha<1$) than the idle units in systen A ($\mu_A=1$), and therefore $F_B<F_A$ always. However, system B is also more efficient binding inhibitors and additionally $\mu_B$ $(>1)$ units are blocked per bound inhibitor.
Therefore, there is a critical value of $\dot{S}$ for which the fraction of blocked units in system B is large enough to reduce its production rate below that of system A.
Figure \ref{fig:3} shows the normalized critical value $\dot{S}_c(\mu_B)/N_e$ of the substrate supply rate at which the production rates of the two systems are the same (i.e. $\dot{P}_A=\dot{P}_B=\dot{S}/(2+\beta))$. For $\dot{S}>\dot{S}_c$ the production rate of system A is larger than that of system B.
For low values of $\mu_B$ the condition $\dot{P}_A=\dot{P}_B$ is reached for high values of $\dot{S}$, and therefore, low values of $F_B$.
For $F_B\ll 1$, eqs. (\ref{stat:7}-\ref{stat:10}) give
\begin{equation}
\frac{\dot{S}_c}{N_e} \approx \frac{2+\beta}{\tau + \mu_B \frac{\tau_I \beta}{2+\beta}}.
\label{stat:11}
\end{equation}
For larger values of $\mu_B$, the critical value $\dot{S}_c$ must be obtained numerically. As shown in Fig. \ref{fig:3}, $\dot{S}_c$ decreases as $\mu_B$, $\beta$ or $\tau_I$ increase or $\alpha$ increases.

\begin{figure}
\resizebox{0.45\textwidth}{!}{%
  \includegraphics{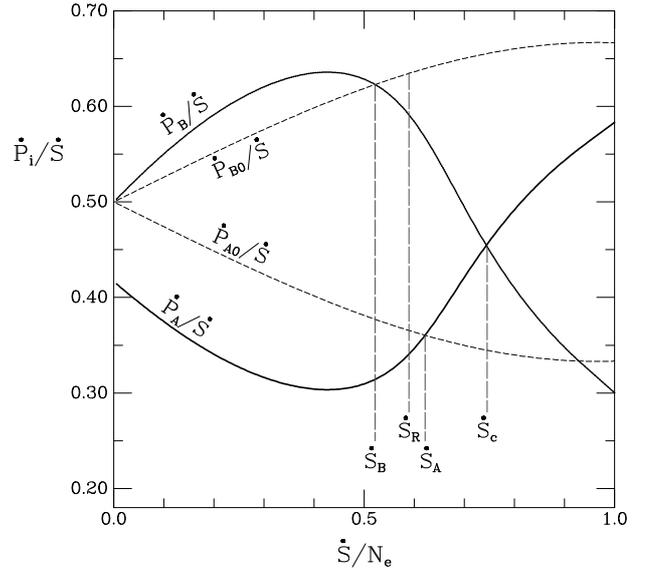}
}
\caption{
Normalized production rates of systems A and B. The continuous curves correspond to the case when system A produces
inhibitors ($\beta_A=0.2$). The dashed curves correspond to the case when there is no inhibitor production.
In the two cases  $p_0=0.1$, $\alpha=0.5$, $\tau_A=\tau_B=1$, $\tau_{I,A}=\tau_{I,B}=5$, $\mu_A=1$ and $\mu_B=4$. The four vertical dashed lines indicate the values $\dot{S}_A$,  $\dot{S}_B$, $\dot{S}_R$, and $\dot{S}_c$ at which 
$\dot{P}_A=\dot{P}_{A,0}$, $\dot{P}_B=\dot{P}_{B,0}$, $\dot{P}_{B,0}/\dot{P}_{A,0}=\dot{P}_{B}/\dot{P}_{A}$, and $\dot{P}_A=\dot{P}_B$, respectively.
}
\label{fig:4}
\end{figure}

In order to illustrate how the distribution of performances changes when inhibitors are released, we now perform a comparison between the previous case and a case where none of the systems release inhibitors. Fig. \ref{fig:4} shows the normalized production rates $\dot{P}_A$ and $\dot{P}_B$ as a function of $\dot{S}/N_e$ when system A converts a fraction $\beta$ of its production into inhibitors (A$\uparrow$\&B$\downarrow$, continuous curves), as well as the production rates $\dot{P}_{A,0}$ and $\dot{P}_{B,0}$ when none of the systems produce inhibitors (A$\downarrow$\&B$\downarrow$, dashed curves). Note that depending on the value of $\dot{S}$, the production of inhibitors by system A can increase the production of system B ($\dot{S}<\dot{S}_B$), can
increase the production of system A ($\dot{S}>\dot{S}_A$), can decrease the ratio $\dot{P}_B/\dot{P}_A$ of
the systems production ($\dot{S}>\dot{S}_R$), or can make the production of system A larger than that of system B ($\dot{S}>\dot{S}_c$). For the case shown in Fig. \ref{fig:4}, $\dot{S}_A=0.62$,  $\dot{S}_B=0.53$, $\dot{S}_R=0.59$, and $\dot{S}_c=0.74$. These values depend on the model parameters: for example, if $\alpha=0.75$ and $\beta=0.2$ then $\dot{S}_A=0.60$,  $\dot{S}_B=0.11$, $\dot{S}_R=0.46$, and $\dot{S}_c=0.61$, 
whereas, if $\alpha=0.5$ and $\beta=0.1$ then $\dot{S}_A=0.75$,  $\dot{S}_B=0.64$, $\dot{S}_R=0.72$, and $\dot{S}_c=1.03$.

The stationary solutions obtained above show that our choice of mechanisms leads to a model that displays some of the distinctive behaviors of a set of systems in competition for a resource, where each system has the possibility to increase its efficiency by intra-system cooperation and/or to to reduce the performance of its competitors by investing part of its profits to inflict aggression. That is,
i) Diversity: Even in the case of identical systems there are stable inhomogeneous solutions.
ii) Hysteresis: in the bi-valuated region, a system steadily producing inhibitors needs large perturbations to leave this regime and viceversa.
iii) Aggression may give comparative advantages: when one or more systems are producing inhibitors, the total
production is always below the maximum possible value. However, the distribution of profits can change in
such a way that a system producing inhibitors can increase its production and, even more, can increase
its ranking in the systems' performance distribution. This comparative advantage is relevant when the production ranking among the systems determines the chances for survival and reproduction.

Note that if we add a third system C, composed by non-cooperative units ($\mu_C=1$ as system A) that never releases inhibitors ($\dot{I}_C=0$ as system B), it will always perform better than system A. This kind of ``opportunism'' is relevant for the coevolution of the system parameters under selection pressure studied in Sec. \ref{sec:4}. 

\subsection{Discrete non stationary model}
\label{subsec:2.2}
We now consider a discrete non stationary model that avoids several approximations adopted in the Mean Field stationary model described in Sec. \ref{subsec:2.1}. Specifically:
1) In the discrete model the systems can have an inhomogeneous internal structure; i.e. a system can be composed of cooperative arrangements of various sizes. 
2) The state of each unit, and therefore the state of its neighbors in a cooperative arrangement, is known at any time. This is a relevant difference with respect to the Mean Field stationary model because the probability of an idle unit in a cooperative arrangement binding to a substrate depends on the number of busy neighbors in the arrangement: once any idle unit binds to a substrate, the other units in the arrangement increase their probability to bind to substrates; therefore some cooperative arrangement tend to be fully busy while others remain fully idle.
In contrast, in the mean field approximation the mean number of busy units 
$\langle\kappa\rangle$ was approximated as $(\mu-1)(1-F)$. This approximation neglects the fact that the inhibition of a cooperative arrangement decreases $F$ but does not increase the probability of an idle unit binding to a substrate.
3) Substrates, products and inhibitors are discrete.
4) A system fulfilling the conditions to release inhibitors at a given time can convert a product into an inhibitor only if a product was produced at that time.

As in the Mean Field stationary model, there is a set of $N_g$ systems, each one possesing the same number $N_e$ of units, and the $N_g \times N_e $ units compete to acquire the available resources that are supplied at a rate $\dot{S}$. However, now the state of each unit is characterized by an integer phase variable $\phi_{i,j}(t)$ where $1\leq i \leq N_g$ indicates the system, $1\leq j \leq N_e$ enumerates the units in the system, and $t$ is a discrete time counter. The phase of unit $(i,j)$ evolves in a similar way as the phase of the stochastic automata originally proposed by Mikhailov and Hess \cite{mikha96,stange98,stange99,mikha02}, but allowing for two additional processes: (i) the unit can bind either to a substrate or to an inhibitor and (ii) units can work in cooperation with other units, thus modifying their ability to bind to a substrate or to an inhibitor. The idle state corresponds to the phase $\phi=0$, where the unit is ready to acquire either a substrate or an inhibitor. When the unit acquires a substrate the phase changes to $\phi=1$; afterwards the phase value is increased in each time step until the unit reaches the maximum phase value $\phi=\tau$; once this happens the unit returns to its original idle state $\phi=0$. A product is released at a fixed phase $1<\tau_p<\tau$. When the unit acquires an inhibitor the phase changes from $\phi=0$ to the negative value $\phi=-\tau_I$; afterwards the phase value is increased by one unit in each time step until the unit reaches its idle state $\phi=0$. The algorithm to iterate the phase of unit $(i,j)$ is then
\begin{small}
\begin{eqnarray}
\lefteqn{\phi_{i,j}(t+1) =}  \nonumber \\
& & \left\{\begin{array}{l@{\quad}l}
\phi_{i,j}(t) + 1 & \rm{if} \; \phi_{i,j}(t) \neq 0,\\
0                 & \rm{if} \;\phi_{i,j}(t) = \tau \\
1                 & \rm{if}\; \phi_{i,j}(t) = 0 \; \rm{with \; probability}\; p_{i,j},\\
-\tau_I           & \rm{if}\; \phi_{i,j}(t) = 0 \; \rm{with \; probability}\; q_{i,j},\\
0                 & \rm{if}\; \phi_{i,j}(t) = 0 \; \rm{with \; probability}\; 1 - p_{i,j} -q_{i,j}.
\end{array} \right. 
\label{eq:1}
\end{eqnarray}
\end{small}
The probabilities $p_{i,j}$ and $q_{i,j}$ are given by
\begin{equation}
p_{i,j}=p_0 N_S/\alpha^{\kappa_{i,j}}\;\;\; {\rm and}\;\;\; q_{i,j}=p_0 N_I/\alpha^{\kappa_{i,j}}.
\label{eq:2}
\end{equation}
During a single iteration $t$ the number of substrates $N_S$ and inhibitors $N_I$ changes as they are bound by idle units. In Eq. (\ref{eq:2}) the parameter $p_0$ represents the probability that an isolated idle unit binds either to a substrate or to an inhibitor when only one substrate or one inhibitor is available to be bound. The parameter $\alpha>0$ controls the degree of cooperation between units and the exponent $\kappa_{i,j}$ is the number of busy units that cooperate with unit $j$ in system $i$. When all the units in a cooperative arrangement are in their idle state, $\kappa_{i,j}=0$ and therefore there is no cooperation; that is, the probability to bind to a substrate is the same as if the units were working in isolation. When $\kappa_{i,j}$ units in a cooperative structure are busy ($1\leq\phi\leq\tau$) the probabilities $p_{i,j}$ and $q_{i,j}$ of the idle units in the structure are increased (if $\alpha<1$) or decreased (if $\alpha>1$) by a factor $\alpha^{-\kappa_{i,j}}$ relative to the isolated case ($\kappa_{i,j}=0$ or $\alpha=1$). When an unit binds an inhibitor, it remains inoperative during $\tau_I$ iterations. If an unit is in its inhibited phase ($-\tau_I\leq\phi\le 0$) the binding probabilities are $p_{i,j}=q_{i,j}=0$ for the idle units that cooperate with the inhibited unit; however the cooperating units that were processing a substrate continue normally their way towards the idle state. In the following we adopt $\alpha<1$.

The configuration of the cooperative units in a system remains fixed during the simulation; that is, if the units $j$ and $j+1$ of system $i$ are set as a cooperative arrangement these two units remain in cooperation during the complete simulation of one generation. For system $i$ the configuration is characterized by the three parameters $M_i$, $D_i$, and $T_i$ giving, respectively, the number of units working in isolation (Monomers), the number of cooperating arrangements of two units (Dimers), and the number of arrangements of four units (Tetramers); all possible configurations satisfy $M_i + 2 D_i + 4 T_i = N_e$. 

At the beginning of a given iteration $\dot{S}$ substrates are added and the following procedure is executed: (a) the phases of the busy and the inhibited units are increased by one step. (b) an unit $(i,j)$ is selected at random among those in their idle state provided it is not associated to an inhibited unit in a cooperative arrangement; (c) the selected unit starts its processing cycle with probability $p_{i,j}$, is inhibited with probability $q_{i,j}$, or remains idle with probability $1-p_{i,j}-q_{i,j}$. If the unit starts its processing cycle the number of substrates $N_S$ decreases by one unit. If an inhibitor is bound, the total number of inhibitors $N_I$ decreases in one unit. If the unit $(i,j)$ does not leave the idle state, it is not selected again during the present iteration. Steps (b) and (c) are repeated until $N_S+N_I=0$, or until all the units that were idle at the beginning of the iteration $t$ are selected.

The number of products released in a given iteration $t$ by the $N_e$ units in system $i$ is denoted as $\dot{P}_i(t)$ and the total production rate as $\dot{P}=\sum_1^{N_g}\dot{P}_i$. The mean production rate per unit and per duty cycle in system $i$ is denoted as $\upsilon_i=\tau\dot{P}_i/N_e$. $\upsilon_i$ is a measure of the performance of system $i$, which in turn depends on the particular configuration of the units within the system. The corresponding mean production rate for all systems is $\upsilon=(1/N_g)\sum_1^{N_g}\upsilon_i=\tau\dot{P}/N_e N_g$. Since the maximum production rate is $\dot{P}=N_e N_g/\tau$, then $0 \leq\upsilon\leq1$. In absence of inhibitors, the maximum production is approached when the binding probability is $p \gg 1/\tau$. If $N_e N_g > \tau \, \dot{S}$ then, the fraction of units in their idle state is $\sim 1-\tau \, \dot{S}/(N_e N_g)$ and in average $\dot{P}=\dot{S}$. However, if the configuration of cooperating arrangements in the $N_g$ systems is inhomogeneous, then the distribution of the production rates $\dot{P}_i$ is also inhomogeneous.

The inhibitors can be introduced by an external source at a rate $\dot{I}_{ext}(t)$ or can be released into the environment by some systems at a rate $\dot{I}_i(t)$. In Secs. \ref{sec:3} and \ref{sec:4} we describe the conditions a system must fulfill to release an inhibitor.

\section{Inhomogeneous configurations}
\label{sec:3}
Consider a set of $N_g=20$ systems, each one consisting on $N_e=120$ units arranged in different cooperative configurations: $M_i=N_e$ for $1\leq i \leq 5$; $D_i=N_e/2$ for $6\leq i \leq 10$; $T_i=N_e/4$ for $11\leq i \leq 15$; and  $M_i=40$, $D_i=20$, and $T_i=10$ for $16\leq i \leq 20$. To quantify the performance of the $i$th system we evaluate the average production rate $\avi$ per unit and per duty cycle $\tau$ in system $i$. Let us label these four types of systems as types M, D, T and MDT and their corresponding production rates as $\avm$, $\avd$, $\avt$  and $\avmdt$. 

\subsection{External release of inhibitors}
\label{subsec:31}
First we consider the case when the set of systems is supplied with a constant rate $\dot{S}$  of substrates and analyse the dependence of $\avm$, $\avd$, $\avt$ and $\avmdt$ on the external rate of release of inhibitors $\dot{I}_{ext}$. The objective of this analysis is to determine how the degree of organization within each system determines its performance in the presence of increasing levels of aggression.

\begin{figure}
\resizebox{0.45\textwidth}{!}{%
  \includegraphics{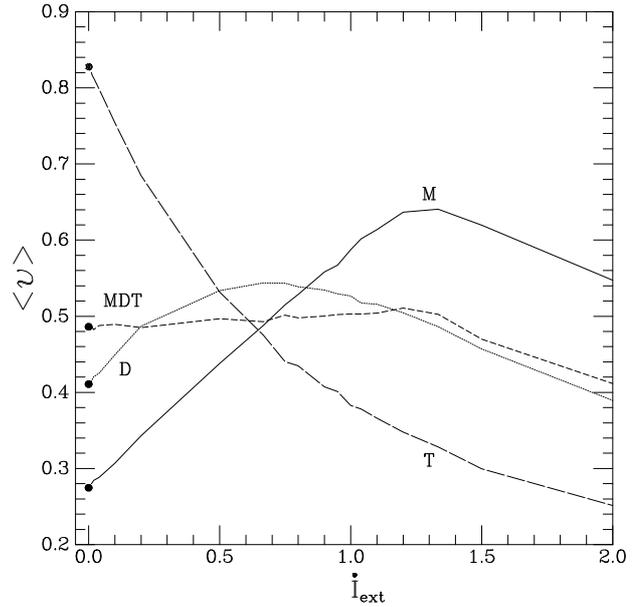}
}
\caption{The average production rate $\av$ per unit and per duty cycle $\tau$ for the four types of systems M, D, T and MDT (see text) as a function of the external rate of release of inhibitors $\dot{I}_{ext}$.
The adopted parameters are $N_e=120$, $\tau=100$, $\tau_I=5\tau$, $p_0=1/\tau$ and $\alpha=1/4$. The substrate supply rate is kept fixed to $\dot{S}=12$. A simulation for $200 \;\tau$ was performed for each value of $\dot{I}_{ext}$.}
\label{fig:5}     
\end{figure}

When $\dot{S}= \frac{1}{2} N_e N_g/\tau$ there are enough substrates to maintain half of the units busy; this corresponds to a production rate $\av=1/2$. The average number of free substrates depends on $p_0$ and on the number of cooperative arrangements in the system. If $p_0=1/\tau$ and $N_S=1$, an isolated unit remains in average half of the time in its idle state. If $p_0$ is reduced, $N_S$ must be increased in the same proportion in order to maintain the units working at half velocity. If not all the units work in isolation, the units in cooperative arrangements will be busier than the isolated ones (for $\alpha<1$). Therefore, in absence of inhibitors one expects $\avt > \avd > \avm$, but these relations change depending on the rate of release of inhibitors $\dot{I}_{ext}$ and on the blocking time $\tau_I$. The results in Fig. \ref{fig:5} correspond to the case when $\dot{S}= \frac{1}{2} N_e N_g/\tau$. As $\dot{I}_{ext}$ increases, the performance of the T-systems decreases whereas for the M-systems it increases: the higher impact of inhibition on the T-systems means that there are less highly-organized units competing for the resource, thereby increasing its availability for the rest of the systems. Following the drug-gang analogy, this means that shootings taking place in busy corners where several dealers operate at the same time force the customers to move to other locations where smaller groups of dealers operate, improving the profits of these dealers. However, for $\dot{I}_{ext}\geq 1.2$ even the performance of the M-systems decreases because more than half of the units are blocked (i.e. $\tau_I \; \dot{I}_{ext} \; \overline{\mu} \geq \frac{1}{2} N_g N_e$, where $\overline{\mu}\simeq 2$ is the mean number of units in cooperating arrangements). This result illustrates how the performance of all systems will be negatively affected regardless of their degree of organization for sufficiently high levels of aggression. Note that the performance of the MDT systems remains almost constant for $\dot{I}_{ext
}< 1.2$. Therefore, at least up to a certain intensity of aggression, diversification in the degree of organization is a robust strategy against aggression.
\begin{figure}
\resizebox{0.47\textwidth}{!}{%
  \includegraphics{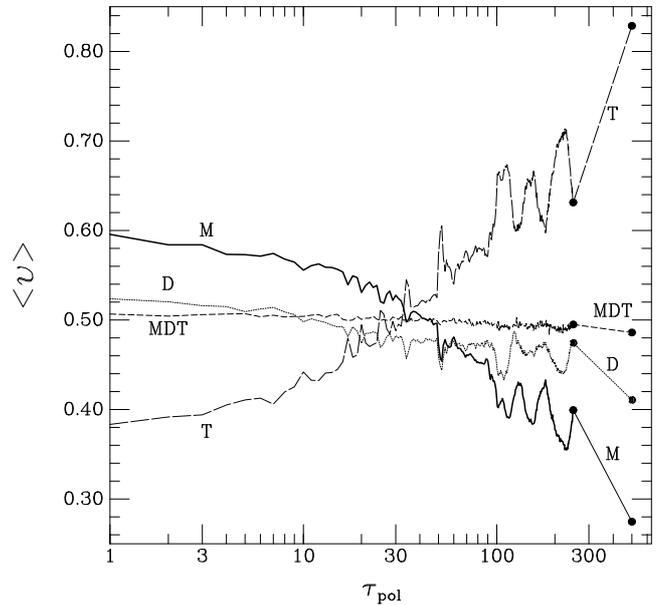}
}
\caption{The average production rates for the four types of systems M, D, T and MDT as a function of the nonbelligerent period $\tau_{non}$ for $<\langle\dot{I}_{ext}\rangle=1$. Model parameters as in Fig. \ref{fig:5}.}
\label{fig:6}     
\end{figure}

In order to examine the effect of a non-stationary inhibitor release rate, Fig. \ref{fig:6} shows the average production rates $\avm$, $\avd$, $\avt$ and $\avmdt$ as a function of the nonbelligerent period $\tau_{non}$. The inhibitor release rate is $\dot{I}_{ext}=0$ during $\tau_{non}-1$ iterations and in the next iteration the release rate is $\dot{I}_{non}=\tau_{non}$; that is, the average inhibitor release rate is $\langle\dot{I}_{ext}\rangle=1$. In order to reduce synchronization effects the inhibition time was randomized 10\% around its mean value $\tau_I=5\tau$. The results show how the sudden onset of aggression affects the systems: those with a greater degree of organization have a better performance as the nonbelligerent period grows longer due to their greater efficiency. Note that as $\tau_{non}$ increases the average production rates tend to the values corresponding to $\dot{I}_{ext}=0$. When a large number of inhibitors are added in a single iteration ($\tau_{non}\gg 1$) many idle units in all cooperating arrangements are blocked, and therefore all types of systems reduce their performances during $\tau_I$ iterations. The performance of the MDT systems remains almost constant for a wide range of values of $\tau_{non}$, illustrating once again the robustness of the systems with diverse degrees of organization.

\subsection{Internal release of inhibitors}
\label{subsec:32}
Consider the case in which inhibitors are produced under certain conditions and at a given cost by the systems. We assume that at a given time an inhibitor is released into the environment by system $i$ if the following conditions are fulfilled: (i) at least one of the units in system $i$ has released a product during the present iteration $t$; (ii) there are no free inhibitors; and (iii) $\vcri > \barvi(t)  > \vcri/2$, where $\barvi$ is the system production averaged during the previous $\tau_{ave}$ iterations and $\vcri$ is the critical average production rate below which a system will release an inhibitor. Since an aggression has an associated cost, a system must achieve a minimum level of performance in order to release an inhibitor; therefore, as in the case of the continuous Mean Field model, an inhibitor will not be released if $\barvi(t)  < \vcri/2$. The cost of producing an inhibitor is one product. During a given iteration a given system may release more than one inhibitor, but in general, conditions (i) and (ii) limit the release of inhibitors to at most to one per iteration. This corresponds to low intensity conflicts in the drug-gang example. Continuing with this analogy, an aggression takes place once the profits of a particular gang fall below a given critical value, however if the profits are too low the gang will not be able to afford an aggression.
\begin{figure}
\resizebox{0.47\textwidth}{!}{%
  \includegraphics{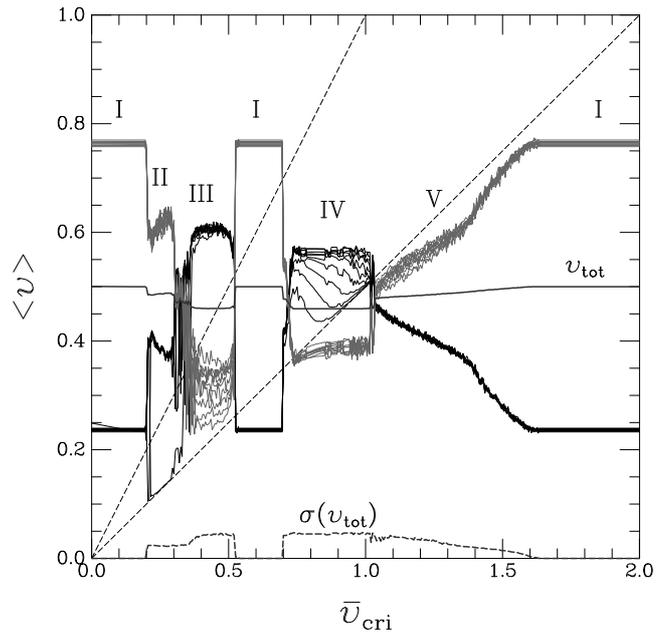}
}
\caption{The average production rate $\avi$ of the 20 competing systems as a function of the critical averaged production $\vcri$ (see text). The model parameters are the same as in Fig. \ref{fig:5} and $\tau_{ave}=10 \;\tau$. Half of the systems (black curves) have their units working in isolation (M-systems) and the other half (grey curves) have their units working in cooperative arrangements of four units (T-systems). The curve labelled $\upsilon_{tot}$ corresponds to the total average production and the curve labelled $\sigma(\upsilon_{tot})$ to its standard deviation. In the region in between the two diagonal dashed lines the condition (iii) is fullfilled. The Roman numbers label the 5 different regimes displayed by the systems.}
\label{fig:7}     
\end{figure}

Fig. \ref{fig:7} shows the average production rates $\avi$ as function of $\vcri$ for 20 systems competing to acquire the substrates supplied at a constant rate $\dot{S}= \frac{1}{2} N_e N_g/\tau$. Half of the systems (black lines) have their units ($N_e=120$) working in isolation (M-systems) and the other half (grey lines) have their units working in cooperative arrangements of four units (T-systems). The curve labelled $\upsilon_{tot}$ corresponds to all-systems average production per unit and the curve labelled $\sigma(\upsilon_{tot})$ gives its standard deviation. In the region in between the diagonal dashed lines the condition (iii) is fullfilled. The Roman numbers label the 5 different regimes observed in Fig. \ref{fig:7}. In regime (I) none of the system production falls in the range [$\vcri/2,\vcri$] and therefore there is no inhibitor production; the cooperation in T-systems results in a better performance than in M-systems. In regime (II) a few M-systems releasing inhibitors (in fact, only one in Fig. \ref{fig:3}) are able to substantially reduce the production of T-systems and to increase the production of the remaining non belligerent M-systems; however, T-systems perform better than M-systems. Note that the system is self-regulated to mantain the production of the attacking system $k$ close to $\overline{\upsilon}_k = \vcri/2$. This is an interesting instance of emergent cooperation: when the value of $\vcri$ is modest, some of the less organized systems keep their performance just above $\vcri/2$, negatively affecting the performance of those systems with a greater degree of organization and thereby improving the availability of substrate for the rest of their non-belligerent, less-organized peers. In regime (III) all T-systems release inhibitors. The performance of T-systems is reduced because of the cost of producing inhibitors and the blocking of their cooperating units. In this regime the M-systems perform the best, but the all-systems average production $\upsilon_{sys} $ is reduced by about 10
 \%.  In regime (IV) all M-systems release inhibitors and their performance is higher than that of T-systems; the disadvantage associated to the cost of releasing inhibitors is over-compensated by the fact that an inhibitor is able to block four units in the T-systems. As in regime (III), the all-system production $\upsilon_{sys}$ is reduced by about 10\%. In regime (V) the T-systems sporadically release inhibitors in such a way that their averaged production rates $\avi$ remains close but below $\vcri/2$ (note that during the simulation $\barvi$ fluctuates around $\avi$ and sporadically $\vcri > \barvi(t)  > \vcri/2$). These results suggest that full-scale, one-sided aggression is profitable only for less-organized systems, as illustrated by regime (IV): regime (III) shows that full-blown aggression from T-systems has a detrimental effect in their own performance.

\section{Evolution of system parameters}
\label{sec:4}
The behavior of the set of systems depends on the arrangement of the units in each system and on the particular conditions at which each system releases inhibitors. The results in Fig. \ref{fig:7} show that small changes in the system parameters can drastically change the distribution of performances. In Fig. \ref{fig:7} we considered that the conditions to release inhibitors were the same for all systems and that half of the systems were homogeneous M-systems and the other half homogeneous T-systems. Now we allow heterogeneity in both parameters; that is, systems are characterized by the parameters ($f_M^i$,$\vcri^i$), where the monomer fraction $f_M^i$ is the fraction of the units working in isolation in system $i$ (the rest works in cooperative arrangements of four units), and $\vcri^i$ is the value of the average production rate below which system $i$ may release inhibitors. Note that in Fig. \ref{fig:7} we used the condition $\vcri > \barvi(t) > \vcri/2$ for the release of inhibitors, but now we use $\barvi(t) < \vcri^i$. For a given set of system parameters ($f_M^i$,$\vcri^i$) there is a distribution of the performances $\avi$ in a simulation. A change of the parameters $f_M^i$ or $\vcri^i$ in one of the systems generally results in a redistribution of the performances $\avi$ and in a change of the all-systems performance $\upsilon_{tot}$.
\begin{figure*}
\includegraphics{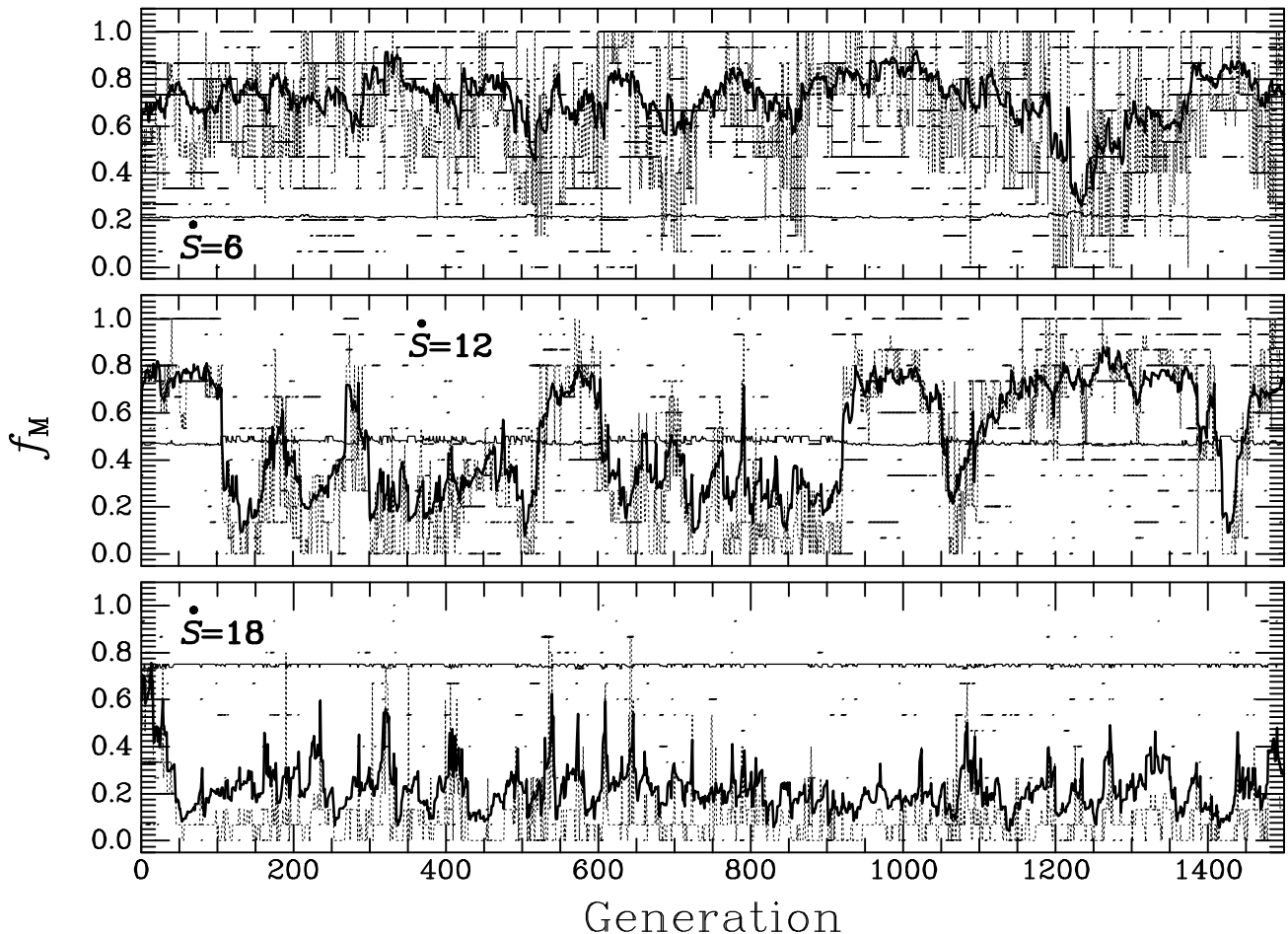}
\vspace*{0.8cm}     
\caption{Evolution of the monomer fraction of the population during 1500 generations for the three labelled values of the substrate supply $\dot{S}$. The thick continuous curve shows the evolution of the monomer fraction $f_M=\sum f_M^i /N_g$ of the population, the dotted curve with large fluctuations shows the monomer fraction of the system with performance ranked at half way $(f_M)_{med}$, the dots indicate the monomer fraction of the systems that have released inhibitors during the simulation of a generation, and the almost horizontal thin curve shows the total average production $\upsilon_{tot}$. The model parameters are $N_g=20$, $N_e=120$, $\tau=100$, 
$\tau_I=6 \;\tau$, $\tau_{ave}=10 \;\tau$, $p_0=1/\tau$, $\alpha=1/4$, $p_{cross} =0.05$, and $p_{mutate}=0.01$.}
\label{fig:8}      
\end{figure*}

A genetic algorithm is implemented to analyse the coevolution of the system parameters ($f_M^i$,$\vcri^i$) with $0\leq f_M^i\leq 1$ and $0\leq \vcri^i\leq 1$. The chromosome of each system is a eight digit binary number; the first four digits give the 16 possible values of the momomer fraction $f_M^i$ ($=0, 1/15, 2/15, \ldots 15/15$) and the last four digits give the critical average production rate $\vcri^i$ below which the system may release inhibitors. For the first generation the parameters ($f_M^i$,$\vcri^i$) are chosen at random, while the parameters $N_g$, $N_e$, $\tau$, $\tau_I$, $\tau_{ave}$, $\alpha$ and the substrate supply rate $\dot{S}$ remain fixed for all generations. Each generation consists of $t_{sim}$ iterations of the model. At the end of a generation simulation the parameters ($f_M^i$,$\vcri^i$) for the next generation (offspring) are set by the following procedure: (i) the performances $\avi$ of the $N_g$ systems are ranked in decreasing order. (ii) those systems whose performances are more than one standard deviation above $\langle\upsilon_{tot}\rangle$ have two offspring and the rest (in decreasing order of performance) have one offspring until the population size $N_g$ is reached. Note that all systems have an offspring unless one or more systems have performances which are more than one standard deviation above $\langle\upsilon_{tot}\rangle$. (iii) Crossover between a random pair of offspring chromosomes occurs with a probability $p_{cross}$, and offspring mutate with a probability $p_{mutate}$ by changing one of their chromosome digits. The system with the best performance is always reproduced  
without any change.

\begin{figure}
\resizebox{0.45\textwidth}{!}{%
  \includegraphics{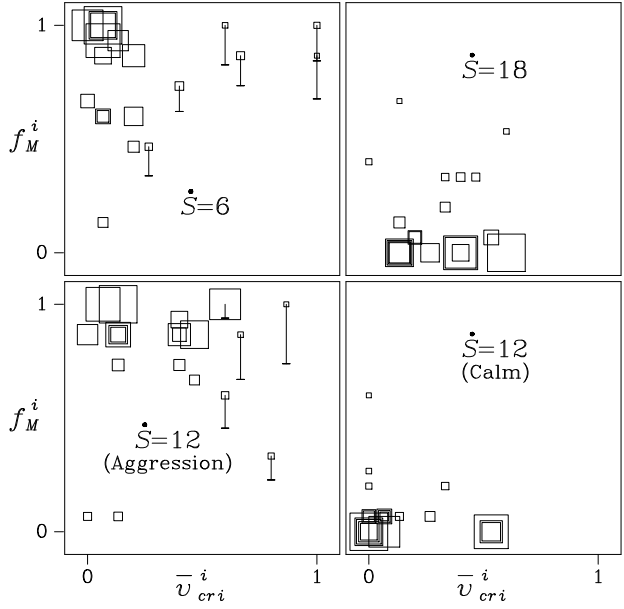}
}
\caption{
Distribution of the parameters $f_M^i$ and $\vcri^i$ for the 20 systems in four particular generations. 
The bottom-left panel corresponds to generation 850 and the rest of the panels to generation 1000.
The labels indicate the value of $\dot{S}$. The size of the square representing each system is proportional to its ranking in the performance distribution, the larger corresponds to the best performer. The vertical line associated to some of the squares indicates that the system has released inhibitors during the simulation of this particular generation; the size of the line is proportional to the amount of inhibitors released.}
\label{fig:9}
\end{figure}

Figure \ref{fig:8} shows the results for three different values of the substrate supply rate ($\dot{S}=1/3$, $1/2$ and $2/3$ of the full occupation value $\dot{S}_F=N_e N_g/\tau$). In each case the evolution for 1500 generations is shown. The thick continuous curve shows the evolution of the mean monomer fraction $f_M=\sum f_M^i /N_g$ of the population, the dotted curve with large fluctuations shows the monomer fraction of the system with performance ranked at half way $(f_M)_{med}$, the dots indicate the monomer fraction of the systems that have released inhibitors during the simulation of a generation, and the almost horizontal thin curve gives the all-systems average production $\upsilon_{tot}$. 
As $\dot{S}$ increases (from top to bottom in Fig. \ref{fig:8}) the average monomer fraction over generations $\barfm$ decreases, the fluctuations of the monomer fraction of the half ranked system $(f_M)_{med}$ decrease, and the number of systems releasing inhibitors decreases. 
Figure \ref{fig:9} shows, for four particular generations, the distribution of the parameters $f_M^i$ and $\vcri^i$ for the 20 systems. Note the diversity of system parameters values in all cases. The distribution of parameters values remains relatively stable for low and high values of $\dot{S}$ (see the top and bottom panels in Fig. \ref{fig:8} and the two top panels of Fig. \ref{fig:9}) but for intermediate values of $\dot{S}$ the distribution of the system parameters values displays two very different distributions (see the center panel in Fig. \ref{fig:8} and the two botton panels of Fig. \ref{fig:9}).

Under conditions of substrate scarcity ($\dot{S}=6$) most of the systems have a large fraction of their units operating in isolation, as the intensity of the aggressions (as reflected by the number of systems releasing inhibitors in the top panel of Fig. \ref{fig:8}) makes collaborative arrays an ill-advised strategy. 
Note that the best performing systems in the upper-left panel of Fig. \ref{fig:9} have high momomer fractions and a low aggression threshold $\vcri^i$.
When the substrate supply is relatively abundant ($\dot{S}=18$) there are few systems releasing inhibitors in each generation, and most of the systems have a large fraction of their units operating in tetrameric configurations; the upper-right panel of Fig. \ref{fig:9} shows that that the best performers are indeed those systems with a low monomer fraction.
When the substrate supply is halfway between abundance and scarcity ($\dot{S}=12$), two quasistable behaviors are observed, as reflected in the central panel of Fig. \ref{fig:8} and the two bottom panels of Fig. \ref{fig:9}: the systems switch between periods of frequent inhibitor release (aggression periods) and high $\barfm$ values, to periods where few inhibitors are released (calm periods) together with low $\barfm$ values (note that $(f_M)_{med}= 0$ for many generations in these quiet periods).

\begin{figure}
\resizebox{0.45\textwidth}{!}{%
  \includegraphics{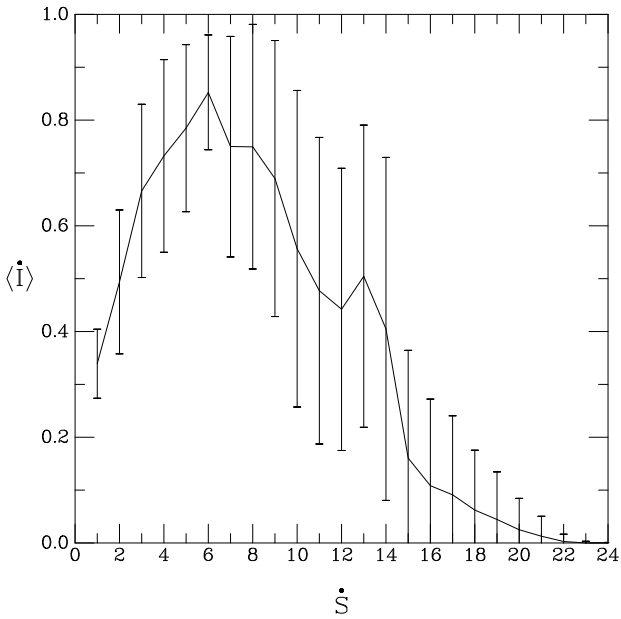}
}
\caption{The average inhibitor release rate $\langle \dot{I} \rangle$ by the set of systems for the last 1400 generations in 5 simulations as a function of the substrate supply rate $\dot{S}$. The error bars correspond to one standard deviation. Model parameters as in Fig. \ref{fig:8}.}
\label{fig10}     
\end{figure}

\begin{figure}
\resizebox{0.45\textwidth}{!}{%
  \includegraphics{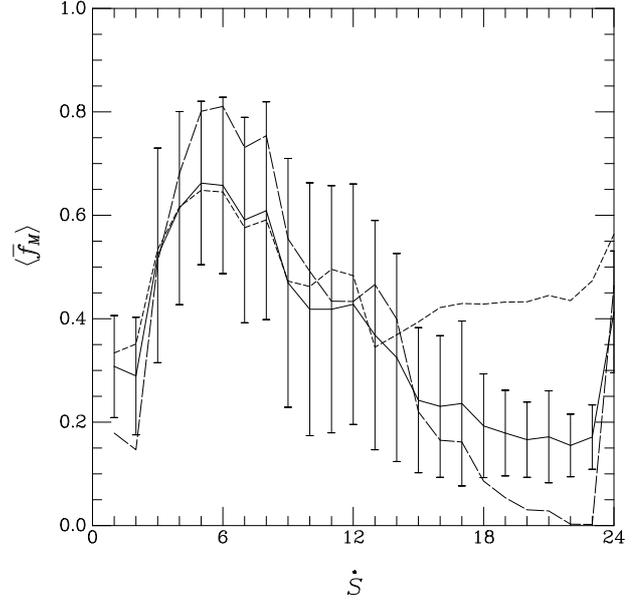}
}
\caption{The average $\langle\barfm\rangle$ of the mean monomer fraction $f_M$ of the population for the last 1400 generations in 5 simulations as a function of the substrate supply rate $\dot{S}$. The long-dash ($\barftop$ vs $\dot{S}$) and the short-dash ($\barfbot$ vs $\dot{S}$) curves show respectively the corresponding averages for the five systems with best and worst performances in each generation. The error bars correspond to one standard deviation. Model parameters as in Fig. \ref{fig:8}.}
\label{fig:11}     
\end{figure}

\begin{figure}
\resizebox{0.45\textwidth}{!}{%
  \includegraphics{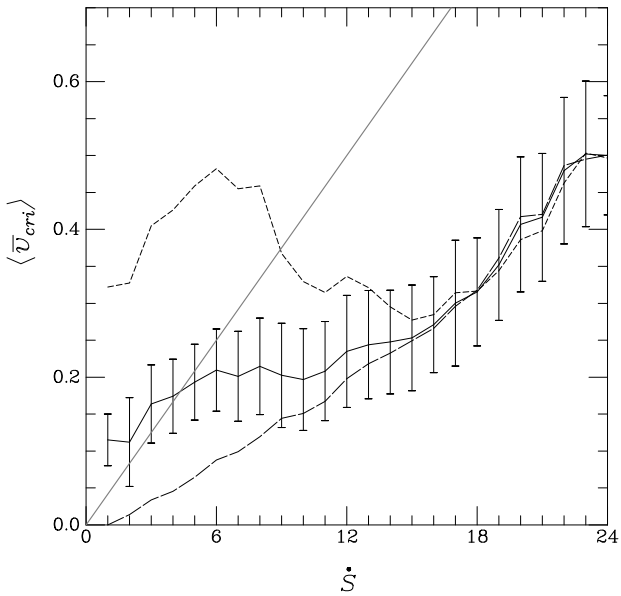}
}
\caption{The average $\langle\overline{\upsilon}_{cri}\rangle$ of $\vcri=\sum \vcri^i /N_g$ in the $5\times 1400$ generations as a function of the substrate supply rate $\dot{S}$. The long-dash ($\langle\overline{\upsilon}_{cri,top5}\rangle$) and the short-dash ($\langle\overline{\upsilon}_{cri,bot5}\rangle$) curves show respectively the corresponding averages for the five systems with best and worst performances in each generation. The grey diagonal line indicates the maximum average production rate per unit and per cycle $\dot{P}_{max}= \dot{S}\tau/N_eN_g$. The error bars correspond to one standard deviation. Model parameters as in Fig. \ref{fig:8}.}
\label{fig:12}     
\end{figure}

In order to examine the general trends as $\dot{S}$ is increased, we performed 5 simulations of 1500 generations for 24 different values of $\dot{S}$ (i.e. $\dot{S}= 1,2,...,24$); the first 100 generations are discarded as well as the first 30$\tau$ iterations in each generation. 
Figure \ref{fig10} shows the average rate of release of inhibitors per iteration and per system $\langle \dot{I} \rangle$ as a function of the substrate supply rate $\dot{S}$.
Figure \ref{fig:11} shows the average monomer fraction in the five simulations, also as a function of the substrate supply rate $\dot{S}$: the continuous curve with error bars corresponds to the average $\barfm$ of the mean monomer fraction $f_M$ of the population, the long-dash curve ($\barftop$ vs $\dot{S}$) corresponds to the same average but taking into account only the five systems with best performances in each generation, while the short-dash curve ($\barfbot$ vs $\dot{S}$) corresponds to the average for the five systems with the worst performances in each generation. 
Figure \ref{fig:12} shows the average $\langle\overline{\upsilon}_{cri}\rangle$ of $\vcri=\sum \vcri^i /N_g$ for each of the 24 substrate supply rate values: the long-dash curve ($\langle\overline{\upsilon}_{cri,top5}\rangle$) and the short-dash curve ($\langle\overline{\upsilon}_{cri,bot5}\rangle$) give respectively the corresponding averages for the five systems with best and worst performances, while the grey diagonal line represents the maximum average production rate per unit and per cycle $\dot{P}_{max} = \dot{S}\tau/N_eN_g$.
The error bars in Figs. \ref{fig10}, \ref{fig:11}, and  \ref{fig:12} correspond to one standard deviation.

It can be observed in Fig. \ref{fig:11} that for $\dot{S} < 2$ the best performers are those systems with a relatively low monomer fraction ($\sim 20\%$). In this situation there are many idle units, and the released inhibitors are not able to reduce the already low system performance because their number is not enough to block a significant fraction of tetramers. Even when many systems release inhibitors in small quantities ($\langle\overline{\upsilon}_{cri}\rangle> \dot{P}_{max}= \dot{S}\tau/N_eN_g$, Fig. \ref{fig:12}), tetrameric configurations are the best strategy to compete for the few available substrates. This type of behavior is also evident in Fig. \ref{fig:4} for $\dot{S} < \dot{S}_C$: in this range, the performance of system B is greater than the performance of system A in spite of the release of inhibitors. For comparison purposes one can assume $N_g=2$, $N_e=10$ and $\tau=1$, so that the full occupation value is $\dot{S}=20$ in that particular case.

For $3 < \dot{S} < 9$ a broad turnover is observed in $\langle \dot{I} \rangle$ and $\langle \barfm \rangle$.
In this range the worst performers release inhibitors since $\langle\overline{\upsilon}_{cri,bot5}\rangle$ is greater than the maximum average production $\dot{P}_{max}$ and $\barfm \approx \barfbot$. The best performers do not release inhibitors (see the top-left panel in Fig. \ref{fig:9}); besides, they have the largest $f_M$ and the lowest  $\upsilon_{cri}$ values. Occasionally, some of the best performers are one standard deviation above the mean performance value; nonetheless, most of the worst performers are reproduced in the next generation. Therefore, in this range most of the systems tend to have a relatively high monomer fraction, with a significant fraction of the systems releasing inhibitors. 
Returning to the drug-gang analogy, when customers are relatively scarce most of the gangs members will operate in solitary, since a greater level of organization would increase their vulnerability; at the same time the gang profits will be reduced both by the use of a less efficient level of organization and by the cost of the aggressions as they battle for customers.

Over $\dot{S}\simeq 9$ most systems have $\vcri^i < \dot{P}_{max}$, thus the number of systems releasing inhibitors decreases as $\dot{S}$ increases, as can be noted in the sharp decline in Fig. \ref{fig10}; therefore, the average monomer fraction $\barfm$ also decreases. 
Note that the largest values of the dispersion of $\barfm$ occurs around $\dot{S}=12$, in agreement with what is observed in the middle panel of Fig. \ref{fig:8}. This high dispersion results from the fact that the system has two quasistable states: one of high monomeric fraction and high inhibitor production (see bottom-left panel in Fig. \ref{fig:9}) , and other of low monomeric fraction and low inhibitor production (see bottom-right panel in Fig. \ref{fig:9}). In the drug-selling gangs example, this situation corresponds to extended periods of relative calm, followed by periods of frequent aggressions. 
For sufficiently high values of $\dot{S}$ the best performers are those systems with a low monomer fraction, showing that when substrate is abundant, inhibitor release is not important and that most of the systems operate using tetrameric configurations (see top-right panel in Fig. \ref{fig:9}).
Finally, the selection pressure on the cooperativity and aggression of systems disappears when the substrate supply rate approaches the full occupation rate $\dot{S}_F=N_e N_g/\tau$.

The results shown correspond to a particular choice of model parameters, rules to generate offspring, conditions to release inhibitors, etc. Nevertheless, these results show that the interplay between cooperative configurations and the possibility to decrease the production of others by the release of inhibitors gives rise to a rich variety of behaviors that are reminiscent of some behaviors observed in real systems that compete for a limited amount of resources. In particular, the model shows that this kind of systems tends to display global evolutive transition when the supply of resources lies in between abundance and scarcity. 

\section{Summary and Conclusions}
\label{sec:5}
We have proposed a model to study the combined effect of competition, cooperation and aggression. The model consists of a set of systems, each one having a number of units that can work with various degrees of organization. The systems compete to acquire the available resources and can attack other systems through the release of inhibitors.

We first considered a Mean Field stationary model in order to examine the main properties of this kind of systems. In particular, this model shows the existence of inhomogeneous solutions (diversity) with the presence of hysteresis. The model also shows how the distribution of performances depends on the degree of organization and the aggression level in the systems. We then introduced a discrete model that allows to follow the state of each unit in each system. The discrete model is first applied to the case where four types of systems are interacting: M, D, T and MDT-systems, where M, D, and T refer to units working as Monomers, Dimers and Tetramers. We have analyzed the production of these four types of coexisting systems as a function of the rate at which inhibitors are released, in order to examine how the degree of organization within each system determines its performance in the presence of aggression and competition. If the release rate of inhibitors is stationary, the T-systems dominate the consumption of the available resources for low release rates whereas at high release rates the M-systems dominate the consumption; however, the performance of every system will be degraded regardless of their degree or organization if the inhibitor release rate is sufficiently intense. Systems with a mixture of degrees of organization exhibited robustness against aggressions, at least up to a given inhibitor release rate. For a non-stationary release of inhibitors the performance of the various types of systems depends on the precise time dependence of the inhibitor release rate. We have shown results for the case when the release of inhibitors occurs sporadically but the time average rate remains constant; in this scenario, those systems with a greater degree of organization performed better; nevertheless those systems with a mixture of several degrees of organization exhibited robustness against aggressions regardless of the specific way in which inhibitors were released.

We have also considered the case in which the systems can release inhibitors into the environment under prefixed conditions and at a given cost. As shown in Fig. \ref{fig:7}, when the systems are allowed to produce inhibitors, distinct regimes occur. In these circumstances, when the cost of an aggression is modest, emergent cooperation may arise between M-systems. We also observed that full-scale, one-sided aggression is profitable only for less organized systems.

Finally, we have allowed for the coevolution of the systems parameters as a function of the substrate supply rate; specifically, we considered the degree of organization within each system (as reflected by the monomer fraction $f_M^i$) and the conditions under which inhibitors are released (as represented by the critical production $\vcri^i$), and have found that the interplay between cooperative configurations and the possibility to decrease the production of others by the release of inhibitors result in a rich variety of behaviors. 
For very low values of the substrate supply rate many units remain idle, and therefore aggressions are irrelevant; in this situation most systems have a high degree of organization.
As the substrate supply rate is increased aggressions become frequent, making systems with a low degree of organization competitive. 
However, when an important fraction of units ($\sim 1/2$) are occupied the set of systems displays global evolutive transitions between periods of calm (low aggression and high degree of organization) and periods of belligerence (high aggression and low degree or organization). In the competing gangs example, this corresponds to extended periods of relative calm, followed by periods of frequent aggressions. 
For larger values of the substrate supply rate the periods of aggression become progressively rarer and shorter,
and finally, when the substrate supply rate approaches the full occupation rate $\dot{S}_F=N_e N_g/\tau$, the selection pressure on the degree of organization and the aggression disappear. 

Among the parameters considered in the model, the number of units in a cooperative arrangement and the cooperativity factor $\alpha$ are particularly relevant, since they determine both the degree of organization and the efficiency of the cooperation between units. The results allow us to infer that further increases in the degree of organization (cooperative arrangements with more than four units) would increase the vulnerability of a system in the presence of aggression (which, as we have seen, is associated to situations of relative resource scarcity), but at the same time they would also increase the productivity of the system when resources are plentiful. We have analyzed a situation where only M-systems and T-systems are involved; if cooperative arrangements of three elements would have been considered instead of T-systems the results would have been analogous, since the degree or organization is still greater than that of M-systems. Computational considerations limited the number of systems $N_g$ as well as the number of units per system $N_e$.
However, we performed simulations for two cases to see if the bistable behavior observed in Sec. \ref{sec:4} for $\dot{S}=\dot{S}_F/2$ persists for different system sizes. We considered the cases $N_e/N_g=60/10$ and $240/40$ with corresponding half-occupation resource supply rates $\dot{S}_F/2 = 3$ and $\dot{S}_F/2 = 48$, respectively. Additionally, we adjust the condition that limits the maximum number of free inhibitors at any time (one in the $N_e/N_g=120/20$ case studied in Sec. \ref{sec:4}) to make these cases comparable. The bistable behavior is also observed in these cases; however as as $N_e$ and $N_g$ are increased the periods of calm and aggression become more stable; that is, they become longer in duration and with less variations around the quasistable states. 

In the simulations shown here the supply of resources and the number and size of systems are held constant for all generations. However, self-regulatory processes can modify the number and size of the systems, and the supply of resources can change in time. In the competing gangs example one expects that the number of costumers per drug-seller evolves in time taking into account the increase of desertions during high violence periods (corresponding to low number of customers per drug-seller) and increase of recruitment during non-violent periods (corresponding to high number of customers per drug-seller). Other questions remain open in this study. How do the trajectories of the model parameters depend on the selection rules and on the mutation rate, on the size and number of systems and on the ambient conditions? What are the spatio-temporal patterns if the model is extended to a spatially structured environment?
However, this model is a first step to understand the interplay between competition, cooperation and aggressions in a general context, and to model the coevolution of the parameters of the interacting systems under selection pressure.

\section*{Acknowledgments}
This work was supported by Consejo de Desarrollo Cient\'ifico, Human\'istico y Tecnol\'ogico of the Universidad de Los Andes, M\'erida,  under grants No. C-1653-09-05-B and I-1118-08-05-B.

\end{document}